\documentclass[pra,onetcolumn,floats,showpacs]{revtex4}
\usepackage{amssymb}
\usepackage{amsmath}
\usepackage{amsthm}

\begin{document}

\title{Quantum discord: "discord" between the whole and its constituent}
\author{Xi-Jun Ren}
\email{renxijun@mail.ustc.edu.cn}
\affiliation{\textit{School of Physics and Electronics, Henan University, Kaifeng 475001,
People's Republic of China}}
\author{Heng Fan}
\email{hfan@iphy.ac.cn}
\affiliation{\textit{Beijing National Laboratory for Condensed Matter Physics, Institute
of Physics, Chinese Academy of Sciences, Beijing 100190, China}}

\begin{abstract}
Quantum discord, a measure of quantum correlation beyond entanglement, is
initially defined as the discord between two classically equivalent while
quantum discordant definitions of mutual information. In this paper, we
report some new interpretations of discord which rely on the differences
between measurement induced effects on the local measured system and the
whole system. Specifically, with proper quantitative definitions introduced
in [Buscemi, Hayashi and Horodecki, Phys. Rev. Lett. 100, 2210504 (2008)],
we find that quantum discord can be interpreted as the differences of
measurement induced disturbance or information gain on the local measured
system and on the whole system. Combined with previous similar results based
on measurement induced entanglement and decoherence, our results provide a
unified view on quantum discord.
\end{abstract}

\maketitle

\section{Introduction}

In quantum theory, quantum systems can have correlations different from
classical correlation. For example, quantum entanglement is a special
quantum correlation which enables quantum teleportation, superdense coding,
quantum key distribution and is considered as an important resource for
quantum computation \cite{horodeckis}. Then, can quantum systems have other
nonclassical correlations beyond entanglement that still provide advantages
over their classical opponents? Ignited by one important algorithm named
deterministic quantum computation with one qubit (DQC1) \cite{knill} which
contains negligible amounts of entanglement during the whole computation
process, much attentions have been paid recently to answer the above
question with different proposed nonclassical correlations. Among these
nonclassical correlations, quantum discord attracts particular attention
\cite{modi}.

The definition of quantum discord \cite{ollivier,henderson} for a bipartite
state $\rho ^{AB}$ comprises two different definitions of quantum mutual
information which are extensions of two equivalent definitions of classical
mutual information. Quantum mutual information,
\begin{equation*}
I^{A:B}(\rho ^{AB})=[S(\rho ^{A})+S(\rho ^{B})-S(\rho ^{AB})]=S(\rho
^{A})-S(A|B),
\end{equation*}%
quantifies the total correlations in a bipartite state $\rho ^{AB}$, where $%
S(A|B)=S(\rho ^{AB})-S(\rho ^{B})$ is the conditional entropy of $A$ side.
An alternative mutual information based on a general measurement $\mathcal{M}%
^{B}$ on party $B$ is,
\begin{equation*}
J_{\mathcal{M}^{B}}^{\leftarrow }(\rho ^{AB})=S(\rho
^{A})-\sum_{m}p(m)S(\rho _{m}^{A})=S(\rho ^{A})-S(A|B_{C}),
\end{equation*}%
where $\rho _{m}^{A}$ corresponds to $A$'s state conditioned on $B$'s
measurement output $m$, $S(A|B_{C})=\sum_{m}p(m)S(\rho _{m}^{A})$ gives $A$%
's averaged entropy conditioned on $B$'s measurement outcomes. Since $J_{%
\mathcal{M}^{B}}^{\leftarrow }(\rho ^{AB})$ is the correlation obtained from
$B$'s local measurement and classical communication from $B$ to $A$, it is
considered as the classical correlation between $A$ and $B$. Subtracting
classical correlation $J_{\mathcal{M}^{B}}^{\leftarrow }(\rho ^{AB})$ from
the total correlation measure $I^{A:B}(\rho ^{AB})$, we obtain a quantum
correlation measure
\begin{equation*}
I^{A:B}(\rho ^{AB})-J_{\mathcal{M}^{B}}^{\leftarrow }(\rho
^{AB})=S(A|B_{C})-S(A|B),
\end{equation*}%
which is measurement dependent. Quantum discord is measurement independent
and is obtained through minimizing the above difference or maximizing the
classical correlation $J_{\mathcal{M}^{B}}^{\leftarrow }(\rho ^{AB})$ over
all possible measurements,%
\begin{eqnarray}
D^{\leftarrow }(\rho ^{AB}) &=&\min_{\mathcal{M}^{B}}[I^{A:B}(\rho ^{AB})-J_{%
\mathcal{M}^{B}}^{\leftarrow }(\rho ^{AB})]  \notag \\
&=&[S(\rho ^{A})+S(\rho ^{B})-S(\rho ^{AB})]-\max_{\mathcal{M}^{B}}[S(\rho
^{A})-\sum_{m}p(m)S(\rho _{m}^{A})]  \notag \\
&=&S(\rho ^{B})-S(\rho ^{AB})+\min_{\mathcal{M}^{B}}\sum_{m}p(m)S(\rho
_{m}^{A})  \notag \\
&=&\min_{\mathcal{M}^{B}}[S(A|B_{C})-S(A|B)].  \label{qd}
\end{eqnarray}%
In order to attain the minimum in Eq.(\ref{qd}), we do not have to go
through all general measurements, set of rank-1 positive operator valued
measurements (POVM) of $B$ is enough. The reason is the concavity of
conditional entropy over the convex set of POVMs and the minimum is attained
on the extremal points of the set of POVMs, which are rank-1 \cite%
{modi,devetak}.

Researches on quantum discord has been developing quickly in recent years,
for a comprehensive and insightful review we recommend Ref.\cite{modi}. In
\cite{datta}, through direct calculations, Datta \textit{et al.} showed that
quantum discord scales with the calculation efficiency which provides the
first quantitative evidence that quantum correlation beyond entanglement
plays a role in the speedup associated with a quantum algorithm. Some
interesting operational interpretations of quantum discord such as in terms
of quantum state merging \cite{horodecki} were put forward \cite%
{cavalcanti,madhok}. In \cite{cornelio}, quantum discord was related with
the irreversibility of entanglement dilution and distillation. The dynamics
of quantum discord were already discussed in \cite{auccaise,streltsov}. In
\cite{streltsov1}, quantum discord was linked to entanglement generation
between the bipartite system and the measuring apparatus. We notice that,
besides entanglement generation, quantum measurement also introduces
disturbance on the measured system and provides information gain which
should have links with quantum discord. In this paper, we link quantum
discord to all these measurement induced effects through their differences
on the measured subsystem and on the whole system, hence giving a unified
view on quantum discord in terms of measurement induced effects.

The organization of this paper is outlined as follows. Firstly, we introduce
the information gain and disturbance of a quantum measurement defined by
Buscemi, Hayashi and Horodecki \cite{buscemi}. These definitions satisfy an
information-disturbance tradeoff relation and balance the information in
quantum measurements. Secondly, we apply them to a local rank-1 POVM
measurement on a bipartite system. It turns out that a general local rank-1
POVM has different disturbances or information gains on the local measured
system and on the whole system. We will show that these differences are
related with quantum discord and provide a unified view on quantum discord
in terms of measurement induced effects. Finally we make our conclusion.

\section{Quantum measurement: quantitative tradeoff between disturbance and
information gain}

Quantum measurement provides information on the measured system, at the same
time it also introduces disturbance and destroys the coherence in the
measured system. The pursuit of proper definitions for information gain and
measurement induced disturbance which satisfy consistent
information-disturbance tradeoff relation takes a long time. It is in \cite%
{buscemi} that, in terms of previous proposed concepts, Buscemi, Hayashi and
Horodecki gave both definitions and obtained the global information balance
for arbitrary measurements. As these definitions are essental for our
discussion, we first give an detailed introduction of them.

A general measurement process $\mathcal{M}^{B}$ on the input system $B$,
described by the input density matrix $\rho ^{B}$ on the
(finite-dimensional) Hilbert space $\mathcal{H}^{B}$, can be described as a
collection of classical outcomes $\mathcal{X}:=\{m\}$, together with a set
of completely positive (CP) maps $\{\mathcal{E}_{m}^{B}\}$, such that, when
the outcome $m$ is observed with probability $p(m)=tr[\mathcal{E}%
_{m}^{B}(\rho ^{B})]$, the corresponding a posteriori state $\rho
_{m}^{B^{\prime }}=\mathcal{E}_{m}^{B}(\rho ^{B})/p(m)$ is output by the
apparatus. Generally speaking, we can think that the action of the
measurement $\mathcal{M}^{B}$ on $\rho ^{B}$\ is given in average by the
mapping
\begin{equation*}
\mathcal{M}^{B}(\rho ^{B}):=\sum_{m}p(m)\rho _{m}^{B^{\prime }}\otimes m^{%
\mathcal{X}}:=\rho ^{B^{\prime }\mathcal{X}},
\end{equation*}%
where $\{\left\vert m^{\mathcal{X}}\right\rangle \}$ is a set of orthonormal
(hence perfectly distinguishable) vectors on the classical register space $%
\mathcal{X}$ of outcomes. Such a measurement process can be realized through
the following indirect measurement model \cite{ozawa}. First, an apparatus $%
Q $ with pure initial state $\phi ^{Q}$ is introduced to interact with $B$
through a suitable unitary interaction $U^{BQ}:BQ\rightarrow B^{\prime
}Q^{\prime }\simeq BQ$. Subsequently, a particular measurement $\mathcal{M}%
^{Q^{\prime }}$, depending also on $U^{BQ}$, is performed on the apparatus $%
Q^{\prime }$. If we further introduce a reference system $R$ purifying the
input state as $\Psi ^{RB}$, $Tr_{R}[\Psi ^{RB}]=\rho ^{B}$, the global
tripartite state after the unitary interaction $U^{BQ}$ is,
\begin{equation*}
\left\vert \Upsilon ^{RB^{\prime }Q^{\prime }}\right\rangle :=(I^{R}\otimes
U^{BQ})(\left\vert \Psi ^{RB}\right\rangle \otimes \left\vert \phi
^{Q}\right\rangle ).
\end{equation*}%
The measurement on the apparatus $Q^{\prime }$ can be chosen such that
\begin{equation}
(I^{RB^{\prime }}\otimes M^{Q^{\prime }})(\Upsilon ^{RB^{\prime }Q^{\prime
}}):=\sum_{m}p(m)\Upsilon _{m}^{RB^{\prime }Q^{\prime \prime }}\otimes m^{%
\mathcal{X}}:=\rho ^{RB^{\prime }Q^{\prime \prime }\mathcal{X}},
\label{indirect}
\end{equation}%
where $\{\Upsilon _{m}^{RB^{\prime }Q^{\prime \prime }}\}$ are pure states
such that
\begin{equation}
Tr_{Q^{\prime \prime }}[\Upsilon _{m}^{RB^{\prime }Q^{\prime \prime
}}]=(I^{R}\otimes \mathcal{E}_{m}^{B})(\Psi ^{RB})/p(m):=\rho
_{m}^{RB^{\prime }},
\end{equation}%
also $Tr_{R}[\rho _{m}^{RB^{\prime }}]=\rho _{m}^{B^{\prime }}$. In this
indirect measurement model, the ancillary apparatus $Q$ helps to disclose
the contributions of inaccessible degrees of freedom to the tradeoff between
information gain and disturbance \cite{buscemi}.

With the indirect measurement model (\ref{indirect}), the information gain $%
\iota (\rho ^{B},\mathcal{M}^{B})$ of measurement $\mathcal{M}^{B}$ on $\rho
^{B}$ is defined in terms of quantum mutual information between reference $R$
and classical outcome $\mathcal{X}$,%
\begin{equation}
\iota (\rho ^{B},\mathcal{M}^{B}):=I^{R:\mathcal{X}}(\rho ^{R\mathcal{X}}),
\label{gain}
\end{equation}%
Eq.(\ref{gain}) shows that the information gain is usually better understood
as being about the remote purifying system $R$, while $B$, correlated with $%
R $, represents just the information carrier that is measured. With the
indirect measurement model (\ref{indirect}), the quantum disturbance
introduced by measurement $\mathcal{M}^{B}$ is defined in terms of coherent
information \cite{schumacher},%
\begin{eqnarray}
\delta (\rho ^{B},\mathcal{M}^{B}) &:&=S(\rho ^{B})-I_{coh}^{R\rightarrow
B^{\prime }\mathcal{X}}(\rho ^{RB^{\prime }\mathcal{X}})  \label{disturbance}
\\
&:&=I^{R:Q^{\prime \prime }\mathcal{X}}(\rho ^{RQ^{\prime \prime }\mathcal{X}%
}),  \notag
\end{eqnarray}%
where coherent information is defined as, $I_{coh}^{A\rightarrow B}(\sigma
^{AB}):=S(\sigma ^{B})-S(\sigma ^{AB})$. Eq.(\ref{disturbance}) shows that
the measurement induced disturbance is equal to the information flow into
both the classical outputs $\mathcal{X}$ and the internal degrees of freedom
of the apparatus or environment $Q^{\prime \prime }$. In \cite{buscemi}, it
is proved that when $\delta (\rho ^{B},\mathcal{M}^{B})$ is infinitely
small, it is always possible to introduce a set of recovering operations $\{%
\mathcal{R}_{m}^{B}\}$ that can asymptotically correct the operations $\{%
\mathcal{E}_{m}^{B}\}$ performed on $B$ by the measurement and recover the
quantum correlations between $R$ and $B$. This confirms the correctness of $%
\delta $ to measure the disturbance. The above two definitions provide a
tradeoff relation between information gain and quantum disturbance of a
quantum measurement,
\begin{equation}
\iota (\rho ^{B},\mathcal{M}^{B})+\Delta (\rho ^{B},\mathcal{M}^{B})=\delta
(\rho ^{B},\mathcal{M}^{B}).  \label{tradeoff}
\end{equation}%
Here, $\Delta (\rho ^{B},\mathcal{M}^{B})=I^{R:Q^{\prime \prime }|\mathcal{X}%
}(\rho ^{RQ^{\prime \prime }\mathcal{X}}),$ measures the missing information
in terms of the hidden correlations between $R$ and internal degrees of
freedom of apparatus or environment which are inaccessible to the observer.

\section{A unified view of quantum discord based on measurement induced
effects}

Now we are ready to apply the above quantitative definitions of measurement
induced disturbance and information gain to a bipartite state to measure its
quantum correlation. For a bipartite state $\rho ^{AB}$ and a local
measurement $\mathcal{M}^{B}$ on $B$, we introduce a reference system $R$
purifying $\rho ^{AB}$ to $\Psi ^{RAB}$ and an ancillary apparatus $Q$ for
the indirect measurement model of $\mathcal{M}^{B}$. Noticing that system $R$
and $A$ purify $\rho ^{B}$, we can directly write down the quantum
disturbance of $\mathcal{M}^{B}$ on $\rho ^{B}$ and on $\rho ^{AB}$
respectively,
\begin{equation}
\delta (\rho ^{B},\mathcal{M}^{B})=S(\rho ^{B})-I_{coh}^{RA\rightarrow
B^{\prime }\mathcal{X}}(\rho ^{RAB^{\prime }\mathcal{X}})=I^{RA:Q^{\prime
\prime }\mathcal{X}}(\rho ^{RAQ^{\prime \prime }\mathcal{X}}),
\label{disturbanceb}
\end{equation}%
\begin{equation}
\delta (\rho ^{AB},\mathcal{M}^{B})=S(\rho ^{AB})-I_{coh}^{R\rightarrow
AB^{\prime }\mathcal{X}}(\rho ^{RAB^{\prime }\mathcal{X}})=I^{R:Q^{\prime
\prime }\mathcal{X}}(\rho ^{RQ^{\prime \prime }\mathcal{X}}).
\label{disturbanceab}
\end{equation}%
Their difference is
\begin{eqnarray}
&&\delta (\rho ^{B},\mathcal{M}^{B})-\delta (\rho ^{AB},M^{B})  \notag \\
&=&S(\rho ^{B})-S(\rho ^{AB})+I_{coh}^{R\rightarrow AB^{\prime }\mathcal{X}%
}(\rho ^{RAB^{\prime }\mathcal{X}})-I_{coh}^{RA\rightarrow B^{\prime }%
\mathcal{X}}(\rho ^{RAB^{\prime }\mathcal{X}})  \notag \\
&=&S(\rho ^{B})-S(\rho ^{AB})+S(\rho ^{AB^{\prime }\mathcal{X}})-S(\rho
^{B^{\prime }\mathcal{X}})  \notag \\
&=&S(\rho ^{B})-S(\rho ^{AB})+\sum_{m}p(m)[S(\rho _{m}^{AB^{\prime
}})-S(\rho _{m}^{B^{\prime }})]  \notag \\
&=&I^{RA:Q^{\prime \prime }\mathcal{X}}(\rho ^{RAQ^{\prime \prime }\mathcal{X%
}})-I^{R:Q^{\prime \prime }\mathcal{X}}(\rho ^{RQ^{\prime \prime }\mathcal{X}%
})  \notag \\
&=&I^{A:Q^{\prime \prime }\mathcal{X|}R}(\rho ^{RAQ^{\prime \prime }\mathcal{%
X}}).  \label{condi-dist}
\end{eqnarray}%
The difference between information gains by measurement $\mathcal{M}^{B}$ on
$B$ and on $AB$ is,%
\begin{eqnarray}
&&\iota (\rho ^{B},\mathcal{M}^{B})-\iota (\rho ^{AB},\mathcal{M}^{B})
\notag \\
&=&I^{RA:\mathcal{X}}(\rho ^{RA\mathcal{X}})-I^{R:\mathcal{X}}(\rho ^{R%
\mathcal{X}})  \notag \\
&=&S(\rho ^{B})-S(\rho ^{AB})+\sum_{m}p(m)[S(\rho _{m}^{R})-S(\rho
_{m}^{RA})]  \notag \\
&=&S(\rho ^{B})-S(\rho ^{AB})+\sum_{m}p(m)[S(\rho _{m}^{AB^{\prime
}Q^{\prime \prime }})-S(\rho _{m}^{B^{\prime }Q^{\prime \prime }})]  \notag
\\
&=&I^{A:\mathcal{X|}R}(\rho ^{RA\mathcal{X}}).  \label{cond-info}
\end{eqnarray}%
Strong subadditivity of von Neumann entropy implies that,
\begin{equation*}
\delta (\rho ^{B},\mathcal{M}^{B})-\delta (\rho ^{AB},\mathcal{M}^{B})\geq
\iota (\rho ^{B},\mathcal{M}^{B})-\iota (\rho ^{AB},\mathcal{M}^{B}),
\end{equation*}%
which also follows from conditional mutual information inequality $%
I^{A:Q^{\prime \prime }\mathcal{X|}R}(\rho ^{RAQ^{\prime \prime }\mathcal{X}%
})\geq I^{A:\mathcal{X|}R}(\rho ^{RA\mathcal{X}})$.

For a general measurement $\mathcal{M}^{B}$,
\begin{equation*}
S(A|B_{C})-S(A|B)\geq \delta (\rho ^{B},\mathcal{M}^{B})-\delta (\rho ^{AB},%
\mathcal{M}^{B}),
\end{equation*}%
which comes from the subadditivity of von Neumann entropy $S(\rho
^{A})+S(\rho ^{B})\geq S(\rho ^{AB})$. However, if $\mathcal{M}^{B}$\ is a
rank-1 POVM, its operator $\mathcal{E}_{m}^{B}$ correlated with outcome $m$
is proportional to a projector. Noticing that $\Upsilon _{m}^{RAB^{\prime
}Q^{\prime \prime }}$ is a pure state \cite{buscemi}, we have $\rho
_{m}^{AB^{\prime }}=\rho _{m}^{A}\otimes \varphi _{m}^{B^{\prime }}$ and
\begin{equation*}
S(A|B_{C})-S(A|B)=\delta (\rho ^{B},\mathcal{M}^{B})-\delta (\rho ^{AB},%
\mathcal{M}^{B}).
\end{equation*}%
Hence, we obtain the following expression for quantum discord in terms of
measurement induced disturbance,%
\begin{equation}
D^{\leftarrow }(\rho ^{AB})=\min_{\mathcal{M}^{B}}[\delta (\rho ^{B},%
\mathcal{M}^{B})-\delta (\rho ^{AB},\mathcal{M}^{B})]  \label{qd-disturbance}
\end{equation}%
where $\mathcal{M}^{B}$ is chosen from rank-1 POVMs.

In addition, if we choose measurement $\mathcal{N}^{B}$ from more restricted
\textquotedblleft good\textquotedblright\ rank-1 POVM set that has zero $%
\Delta (\rho ^{B},\mathcal{N}^{B})$ in tradeoff relation (\ref{tradeoff}),
then we have,
\begin{equation*}
\iota (\rho ^{B},\mathcal{N}^{B})-\iota (\rho ^{AB},\mathcal{N}^{B})=\delta
(\rho ^{B},\mathcal{N}^{B})-\delta (\rho ^{AB},\mathcal{N}%
^{B})=S(A|B_{C})-S(A|B).
\end{equation*}%
For such kind of measurements, information gain is balanced with quantum
disturbance. One kind of such measurement is \textquotedblleft
Single-Kraus\textquotedblright\ or \textquotedblleft multiplicity
free\textquotedblright\ measurement with output states $\rho
_{m}^{AB^{\prime }Q^{\prime \prime }}=\rho _{m}^{A}\otimes \varphi
_{m}^{B^{\prime }}\otimes \omega _{m}^{Q^{\prime \prime }}$ \cite{buscemi}.
Therefore, in terms of measurement information gain, we obtain another
expression for quantum discord,

\begin{equation}
D^{\leftarrow }(\rho ^{AB})=\min_{\mathcal{N}^{B}}[\iota (\rho ^{B},\mathcal{%
N}^{B})-\iota (\rho ^{AB},\mathcal{N}^{B})].  \label{qd-gain}
\end{equation}

Now we make some physical discussions on the above results. When $A$ is only
classically correlated with $B$, it is reasonable to expect that measurement
$\mathcal{M}^{B}$ on $B$ induces equal quantum disturbance on $B$ locally
and on $AB$ as a whole, since $A$ does not contribute to quantum coherence
of $B$, no quantum disturbance either. However, when $A$ and $B$ have
quantum correlations, the situation is different. To be explicit, let us
assume $R$ to be a\ reference system purifying $A$ and $B$. In terms of
quantum coherence, $S(\rho ^{B})$ quantifies the coherent interrelations
between $B$ and $AR$, similarly, $S(\rho ^{AB})$ quantifies the coherent
interrelations between $AB$ and $R$. A measurement on $B$ introduces
disturbance on both of them and we may say quantum coherence of $\delta
(\rho ^{B},\mathcal{M}^{B})$ has been destroyed for $B$, at the same time
quantum coherence of $\delta (\rho ^{AB},\mathcal{M}^{B})$ has been
destroyed for $AB$. When $A$ and $B$ are quantum mechanically correlated, $A$
shares part of $B$'s coherent relations, this part of quantum coherence
certainly experiences the quantum disturbance introduced by the measurement
on $B$, however, it does not exist in $S(\rho ^{AB})$ and its disturbance
naturally will not come up in $\delta (\rho ^{AB},\mathcal{M}^{B})$. In
other words, for a general measurement $\mathcal{M}^{B}$ on $B$, its quantum
disturbance $S(A|B_{C})-S(A|B)$ exists in $S(\rho ^{B})$ but not in $S(\rho
^{AB})$, therefore it is contained in $\delta (\rho ^{B},\mathcal{M}^{B})$
but not in $\delta (\rho ^{AB},\mathcal{M}^{B})$. Similarly, for a good
measurement $\mathcal{N}^{B}$ on $B$, there are information gain $%
S(A|B_{C})-S(A|B)$ in $S(\rho ^{B})$ but not in $S(\rho ^{AB})$. Eq.(\ref%
{qd-disturbance}) and Eq.(\ref{qd-gain}) show that the quantum correlation
quantified by quantum discord can be directly understood as the discrepancy
of measurement induced disturbance and information gain between the local
measured subsystem and the whole system.

In fact, besides quantum disturbance and information gain, there is one
other important effect of measurement, entanglement induced between the
measuring apparatus and the system. For a bipartite system state $\rho ^{AB}$%
, a general measurement $\mathcal{M}^{B}$ on $B$ induces entanglement
between the measuring apparatus $M$ and $B$. Furthermore, if $A$ and $B$
have quantum correlations, the distillable entanglement between $M$ and $B$
is different from the distillable entanglement between $M$ and $AB$, their
minimal discrepancy is equal to quantum discord \cite{streltsov1},
\begin{equation}
D^{\leftarrow }(\rho ^{AB})=\min_{\mathcal{M}^{B}}[E_{D}^{M|AB}-E_{D}^{M|B}],
\label{qd-entangle}
\end{equation}%
where $E_{D}$ is the distillable entanglement.

Restricting measurements to rank-1 projective measurements, we will show
that all the above three expressions of quantum discord (\ref{qd-disturbance}%
,\ref{qd-gain},\ref{qd-entangle}) become equivalent. This point can be made
clear with the following relation between conditional information and
relative entropy,
\begin{equation*}
S(A|B_{C})-S(A|B)=D(\rho ^{AB}||\sum_{m}\Pi _{m}^{B}\rho ^{AB}\Pi
_{m}^{B})-D(\rho ^{B}||\sum_{m}\Pi _{m}^{B}\rho ^{B}\Pi _{m}^{B}),
\end{equation*}%
where $D(\rho ||\sigma )=tr(\rho \ln \rho )-tr(\rho \ln \sigma )$ is the
relative entropy and $D(\rho ^{AB}||\sum_{m}\Pi _{m}^{B}\rho ^{AB}\Pi
_{m}^{B})$, $D(\rho ^{B}||\sum_{m}\Pi _{m}^{B}\rho ^{B}\Pi _{m}^{B})$
correspond to the distillable entanglement $E_{D}^{M|AB}$ and $E_{D}^{M|B}$
respectively \cite{streltsov1}. The definition of quantum disturbance (\ref%
{disturbance}) is in terms of coherent information which is closely related
with decoherence. Therefore, it is possible to relate decoherence to quantum
discord. In \cite{coles}, for rank-1 projective measurements, Coles
discusses the relation between quantum discord and decoherence through the
following relation
\begin{equation*}
D(\rho ^{AB}||\sum_{m}\Pi _{m}^{B}\rho ^{AB}\Pi _{m}^{B})=S(\mathcal{X}|R),
\end{equation*}%
where $R$ is the purifying system of $\rho ^{AB}$ and conditional entropy $S(%
\mathcal{X}|R)$ quantifies the missing information from the purifying system
$R$ which results in the decoherence of measurement $\{\Pi _{m}^{B}\}$.
Furthermore, the equivalence between decoherence and information gain for
rank-1 projective measurement can be found through the following relation,
\begin{equation*}
S(\mathcal{X}|R)-S(\mathcal{X}|RA)=I^{RA:\mathcal{X}}(\rho ^{RA\mathcal{X}%
})-I^{R:\mathcal{X}}(\rho ^{R\mathcal{X}}).
\end{equation*}

In summary, restricting to rank-1 projective measurements $\mathcal{M}%
^{B}=\{\Pi _{m}^{B}\}$, we obtain the following equivalent expressions for
quantum discord with different physical meanings,%
\begin{eqnarray}
D_{\mathcal{M}^{B}=\{\Pi _{m}^{B}\}}^{\leftarrow }(\rho ^{AB}) &=&\min_{%
\mathcal{M}^{B}=\{\Pi _{m}^{B}\}}[\delta (\rho ^{B},\mathcal{M}^{B})-\delta
(\rho ^{AB},\mathcal{M}^{B})]~~~ \\
&=&\min_{\mathcal{M}^{B}=\{\Pi _{m}^{B}\}}[\iota (\rho ^{B},\mathcal{M}%
^{B})-\iota (\rho ^{AB},\mathcal{M}^{B})] \\
&=&\min_{\mathcal{M}^{B}=\{\Pi _{m}^{B}\}}[E_{D}^{M|AB}-E_{D}^{M|B}] \\
&=&\min_{\mathcal{M}^{B}=\{\Pi _{m}^{B}\}}[S(\mathcal{X}|R)-S(\mathcal{X}%
|RA)].
\end{eqnarray}

For rank-1 projective measurements, our results coincide with the results
given in \cite{coles}. However, it should be pointed out that our results
also apply to rank-1 POVMs which are not covered in \cite{coles} but are
needed for optimization of quantum discord \cite{modi}. It is interesting to
note that, for the initial "discord" which is defined as the difference
between two discordant definitions of mutual information in quantum case,
the above four equations provide different physical meanings for "discord"
in terms of the difference between the whole and its constituents.

\section{Conclusion}

In this paper, we discuss the quantum correlation in terms of measurement
induced effects. The differences between measurement effects on the local
measured subsystem and on the whole system are used to measure quantum
correlation. It is shown that for rank-1 POVMs on one subsystem of a
bipartite system, the minimal difference between the measurement induced
disturbance on the measured subsystem and on the whole system corresponds to
quantum discord of the bipartite state. Similarly, minimized difference
between the information gain of the measured subsystem and the whole system
over good rank-1 POVMs also corresponds to quantum discord. Combined with
similar results in terms of measurement induced entanglement and
decoherence, our results provide a unified view on quantum discord in terms
of measurement induced effects.

\emph{Acknowledgments.}--- This work is supported by \textquotedblleft
973\textquotedblright\ program (2010CB922904) and NSFC (10974247, 11175248,
11047174). X-J Ren acknowledges financial support from the education
department of Henan province.

\end{document}